# Design and Development of a Localized E-Commerce Solution for Students focussing on Economical Sharing

Faiz Ahmed, Nitin Kumar Jha, Md Faizan
Guided by Dr. Chitra K.
School of Electronics Engineering
Vellore Institute of Technology, Chennai, India

## Abstract

The rapid adoption of e-commerce has transformed how students access goods and resources. However, existing platforms often fail to address the specific needs of campus communities, where students face challenges such as financial constraints, lack of access to affordable goods, and inefficient resource circulation. This research proposes ShareSpace, a localized web application designed specifically for college students to facilitate the buying, and selling of mainly second-hand goods. By addressing imbalances like surplus items left behind by seniors and shortages experienced by juniors, ShareSpace promotes sustainability and affordability within the campus ecosystem. Leveraging modern technologies such as Node.js, React.js, and MongoDB, the project demonstrates the feasibility of creating a student-centric e-commerce solution. The study highlights how ShareSpace solves the challenges of economical pricing and content moderation using proposed solutions. This study also explores the limitations of existing solutions and evaluates the potential of ShareSpace to encourage sustainable consumption and resourcefulness among students.

**Keywords:** Second Hand Trading Platform, Buying, Selling, E-commerce, Online shopping.

## Introduction

The advent of e-commerce has revolutionized how individuals engage in buying, selling, and exchanging goods, offering unprecedented convenience and accessibility. Despite the growth of global e-commerce platforms like eBay, Amazon, and OLX, these solutions often fail to address the specific needs of niche communities such as college students. College campuses differ from the global community where students face unique challenges: financial constraints, the need for economical goods, and a lack of platforms tailored to their local and immediate requirements. Furthermore, the frequent turnover of students leads to an imbalance — seniors leave behind surplus items, while juniors typically lack access to affordable resources.

This research explores the development of ShareSpace, a localized web application designed to empower college students by facilitating the buying, and selling of goods at economical prices. Unlike general-purpose e-commerce platforms, ShareSpace focuses on fostering sustainability and collaboration within campus communities. By encouraging resource circulation and promoting shared consumption, the platform aims to encourage sustainability.

The study aims to bridge the gap in existing solutions by creating a platform that integrates user-friendly design, secure functionality, and a community-driven approach. ShareSpace not only fills these gaps but also addresses problems in existing solutions like high prices and manual content moderation.

In this paper, we examine the limitations of existing e-commerce platforms, analyse user behaviour and preferences, and present the design and implementation of ShareSpace. By understanding the specific challenges faced by students and leveraging modern technologies, the proposed solution aims to make campus commerce sustainable and community-centric.

# 1. Related Work

This section reviews research literature on resource-sharing platforms and analyses existing software solutions like OLX, Facebook Marketplace, etc. We analyse their strengths and weaknesses to demonstrate the need for a student-focused platform like ShareSpace.

## Literature Review

The idea of resource-sharing platforms has been widely studied in research literature. This section explores previous research related to second hand sharing platforms and their feasibility, community centred e-commerce platforms and user attitude towards general e-commerce platforms.

### Traditional E-commerce Platforms

Wang's research on major e-commerce platforms (Amazon, eBay, Tmall, JD.com, Pinduoduo, and Taobao) reveals distinct consumer preferences based on product categories and price points. The study identifies clear platform preferences among users - for instance, JD.com is preferred for electronic products, while Taobao and Pinduoduo are favoured for low-priced daily necessities. Platform success is significantly influenced by review systems, with positive reviews driving sales and negative reviews having a detrimental impact on consumer trust and purchasing decisions.

### Student-Specific E-commerce Solutions

More and colleagues highlight the unique position of college students as a significant consumer group in the e-commerce industry. Their research emphasizes the need for specialized platforms catering to academic requirements, particularly for textbooks, course materials, and educational technology. This aligns with findings from Miah et al., who specifically studied textbook trading platforms, noting that rising textbook costs create significant financial pressure on students.

### Campus-Based Second-Hand Trading

Several studies focus on campus-specific second-hand trading solutions. Giri et al. present a campus-based application system that facilitates information browsing, online communication, and background data management. Their research emphasizes both environmental and economic benefits, particularly in reducing e-waste, while providing cost-effective solutions for students.

Wei et al.'s research on college students' second-hand trading platforms identifies the significant problem of idle items on campus and proposes solutions to increase item circulation rates. Their work emphasizes the importance of sustainable consumption concepts and environmental protection within academic communities.

## Consumer Behaviour in Online Academic Marketplaces

Novgorodtseva et al.'s study of student consumer strategies in online marketplaces reveals that young people are the most dynamic social group in terms of internet market adoption. Key findings include:

- Students frequently purchase categories such as tickets, music, software, and food delivery online
- Purchase frequency typically ranges from once every six months to once every 1-3 months
- Students particularly value services like internet catalogues and preliminary ordering options

The literature reveals a clear trend toward specialized e-commerce solutions for academic communities, with particular emphasis on second-hand trading and textbook exchange. These platforms serve multiple purposes:

1. Addressing student financial constraints
2. Promoting sustainable consumption
3. Facilitating resource circulation within campus communities
4. Meeting specific academic needs through specialized marketplaces

## Research Gaps

- Long-term sustainability metrics for campus-based trading platforms
- Integration of AI and machine learning for better product matching
- Lack of effective content moderation measures
- Cross-campus platform standardization and scaling
- Security and trust mechanisms in closed academic marketplace environments

# Existing Solutions

### eBay

**Features**:

- A global e-commerce platform supporting auctions and fixed-price sales.
- Offers buyer and seller protection programs to ensure secure transactions.
- Advanced search and filtering options for product discovery.
- Wide range of product categories, including new and second-hand items.
- Integration with PayPal and other payment methods for secure payments.

**Challenges**:

- High listing fees and final value fees may deter sellers, especially for low-cost items.

- Limited community-focused features, making it less appealing for local or campus-level exchanges.
- Complex interface that may not be user-friendly for casual users.

## OLX

**Features**:

- A classifieds platform designed for buying and selling locally.
- Free-to-use with no mandatory commission fees for most transactions.
- Simple interface with direct communication between buyers and sellers via chat.
- Strong presence in many regions with a focus on localized services.

**Challenges**:

- Lacks robust moderation tools, leading to scams or fraudulent listings.
- No specific incentives for affordable or sustainable sharing practices.

## Facebook Marketplace

**Features**:

- Integrated with Facebook profiles, providing social validation for buyers and sellers.
- Supports local sales with features to find items near the user's location.
- Easy-to-use interface accessible directly within the Facebook app.
- No fees for basic transactions, making it cost-effective for casual sellers.

**Challenges**:

- Lack of structured reputation or incentive systems for sellers.
- Limited support for product compliance or price competitiveness checks.

| Feature | OLX | eBay | Facebook Marketplace | ShareSpace |
|---|---|---|---|---|
| User-Generated Listings | Users created listings for local buying and selling | Users create listings in both auction-style and fixed-price formats | Users create listings integrated with their Facebook profiles | Student created listings for resource sharing within campus |
| Search and Filter Options | Basic search and filter options for easy browsing | Advanced search and filtering options across a wide range of categories | Basic search with filters and location-based searches | Simple search functionality with keyword based search and filtering options |
| User Profiles and Ratings | Basic user ratings to build trust | Detailed feedback system for buyers and sellers | Item ratings and user reviews; connected to Facebook profiles | Reputation points for users to encourage economical selling |

| Payment and Delivery Options | Supports various payment methods; delivery options vary by region | Multiple payment methods; shipping and tracking available | Integrated with Facebook Pay for transactions | Hand-to-hand transfer of payment and merchandise |
|---|---|---|---|---|
| Spam Report and Block | Tools to report and block spammy or fraudulent users | Buyer and seller protections through eBay Money Back Guarantee | Security and privacy features with safe buying and selling alerts | Detection and removal of non-compliant content using AI |
| Unique Features | Localized buying and selling, fraud prevention tools | Auction format, Global reach, eBay Money Back Guarantee | Social network integration, Ease of sharing among Facebook friends | Point based reputation system, AI based malicious content detection, Campus focused model |

These existing solutions, while effective in their domains, lack the tailored approach needed for college students, such as localized campus-level focus, incentives for economical pricing, and product listing moderations inline with university norms. ShareSpace addresses these gaps by creating a platform specifically designed to meet the unique needs of student communities.

# 2. Methodology

## Overview

ShareSpace is a web application designed to facilitate the sharing and selling of second-hand goods among college students. It prioritizes economical sharing of second hand goods rather than selling of new ones. Focuses on a local community driven approach within the campus community and hand-to-hand transactions.

## System Design

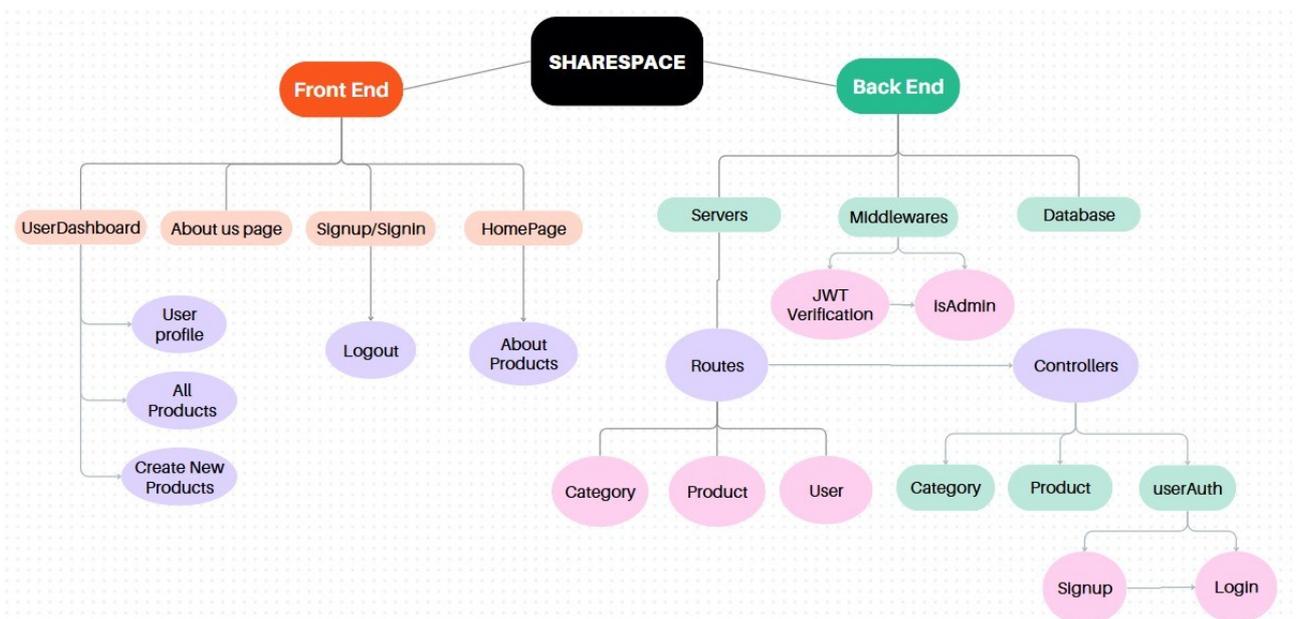

The ShareSpace web application is built upon a client-server architecture. Students interact with the client to perform all their actions, surfing for products, listing and buying products, interacting with their user profiles and so on. The server is responsible for all the business logic, verification, validation, authorization, and other backend tasks.

## Frontend:

**HTML and CSS**: For basic Web Design, layout, and styling.

**Tailwind CSS**: A CSS framework for frontend styling and UI/UX. Used for its responsiveness, ease-of-use, small footprint and interoperability. Enables rapid prototyping and mobile-friendly user interface.

**React JS**: A JavaScript library for building interactive user interfaces, enabling a dynamic experience for users.

## Backend:

**Node.js**: A JavaScript runtime environment that handles asynchronous connections and server API interactions very well.

**Flask (Python)**: A lightweight framework used for implementing business logic and handling API requests to external services, ensuring modularity and flexibility. Ensure easy connectivity with external services due to widely available SDKs and libraries.

**JWT**: JSON Web Tokens used for authentication. Minimal overhead, ability to store data within them and easy-of-use make them a perfect fit.

## Database:

**MongoDB**: A NoSQL document based database software. Offers flexibility in handling unstructured data and file storage, which are important due to early stage of platform rollout and possibility of future changes in college regulations.

## External APIs:

**Google Generative AI SDK**: To connect with Google Generative AI servers for AI integration to verify product listing compliance with platform regulations. Advanced and well-trained GPTs increase the odds of detecting disguised items trying to fool traditional detection systems.

# Core Functionalities

## User Authentication

- The platform securely handles authentication, allowing for user registration and login. The Node.js backend handles the logic, integrating JWT for session management.
- User details are stored in a MongoDB database during registration and accessed on every login.
- Passwords are hashed using the bcrypt hashing function and stored. This is done as a security best practice, as storing plain text passwords is against modern web standards to maintain security and privacy.

## Product Listing

- Users can list their products from their dashboard by submitting product details via a form.
- Form details are validated by ReactJS in the frontend and by Node.js in the backend. This ensures quickest validation locally and also checks for any tampering with the request after frontend validation.
- Compliance check API utilizing word-based blacklist approach and generative AI is used to check compliance with platform's Terms of Service.
- Data is stored in MongoDB, containing various attributes like item name, description, photo, category, and price along with user identifier.

## Product Requests

- Desired products can be requested from the buyers.
- On request, the product is removed from the global pool of products and added to the buyer's local pool of requested products. It stays there until either the transaction is complete or declined.
- The seller receives the product request and both buyer and seller are displayed each other's contact details to coordinate the transaction in person.
- After transaction is completed, the seller marks it as so and the product is deleted.

# Database Design

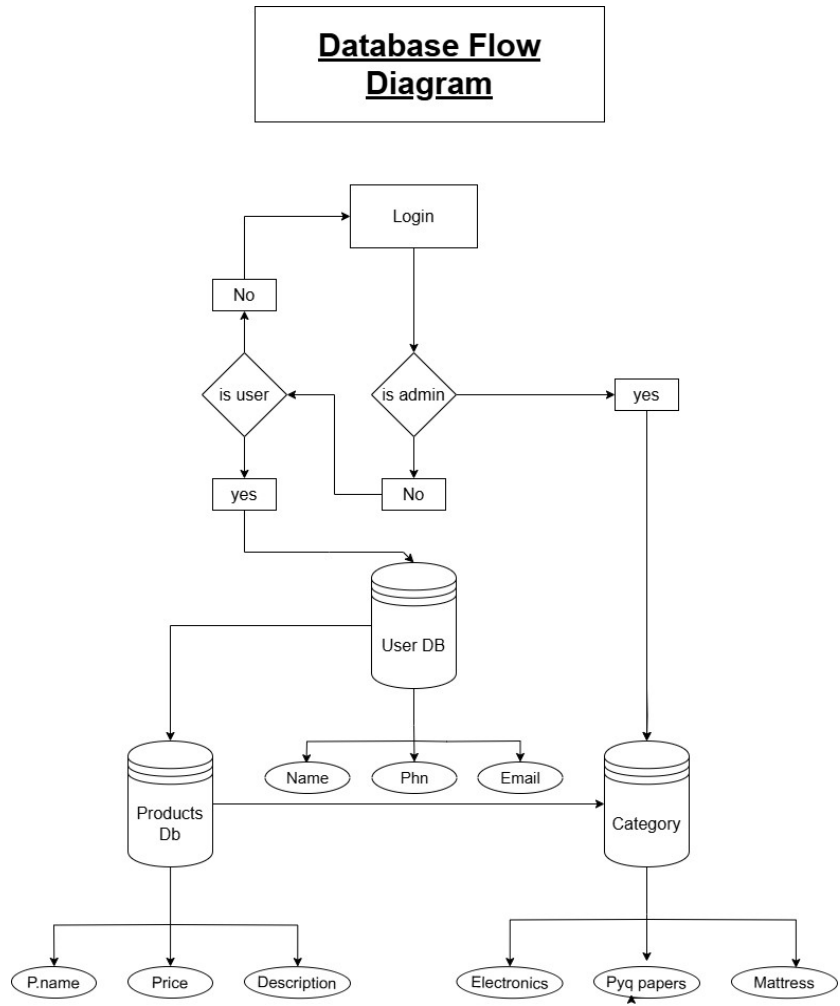

The database consists of 4 main collections:

## 1. Categories Collection

```
{
  _id: ObjectId,
  name: String,    // e.g., "Calculator", "Books", "Laptop", "Mattress"
  __v: Number      // Version key
}
```

## 2. Products Collection

```
{
  _id: ObjectId,
  name: String,
  description: String,
  price: Number,
```

```
    userId: ObjectId,    // Reference to users collection
    category: ObjectId,  // Reference to categories collection
    quantity: Number,
    photo: Object,
    shipping: Boolean,
    createdAt: Date,
    updatedAt: Date,
    __v: Number
}
```

## 3. Users Collection

```
{
    _id: ObjectId,
    name: String,
    email: String,
    password: String,   // Hashed password
    phone: String,
    collegeId: String,
    role: Number,      // 0 for regular users, 1 for admin
    createdAt: Date,
    updatedAt: Date,
    __v: Number
}
```

## 4. OTPs Collection

```
{
    _id: ObjectId,
    email: String,
    otp: String,
    createdAt: Date,
    __v: Number
}
```

# User Workflow

## Account Registration and Login

- Users navigate to the login page and register with their university email id and custom password. On proceeding, a One Time Password (OTP) is sent to their email to verify their student status and email ownership.

- After successful login, a JWT token is created and attached to their session. The user details are stored in the database.

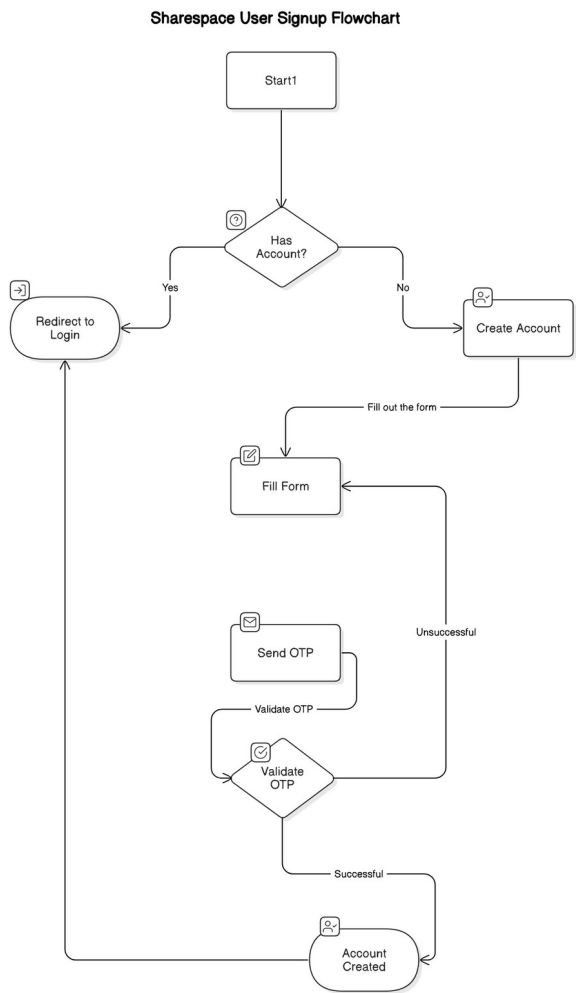
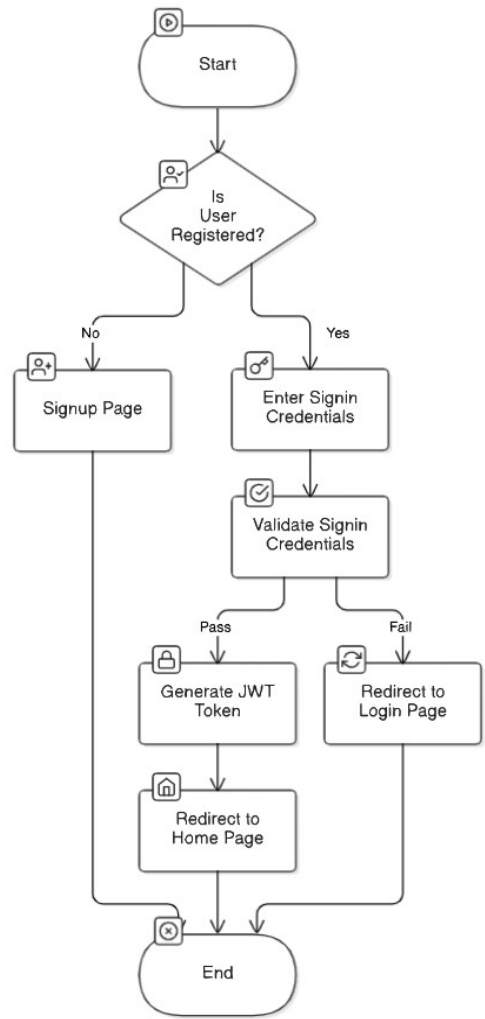

## Product Listing

- Users can list items from their dashboard. They fill the Add Product form with details of the product. Details include a photo of the product, name, description, category and an ideally low price.

- On submitting, the details of the product are checked and verified to be inline with platform's terms of service using AI classification.

- If found to be compliant, price of the product is analysed for is competitiveness by comparing with online sources for similar products.

- Post checking and validation the product is added to the users listings and added to the database.

**Item Buying/Request**

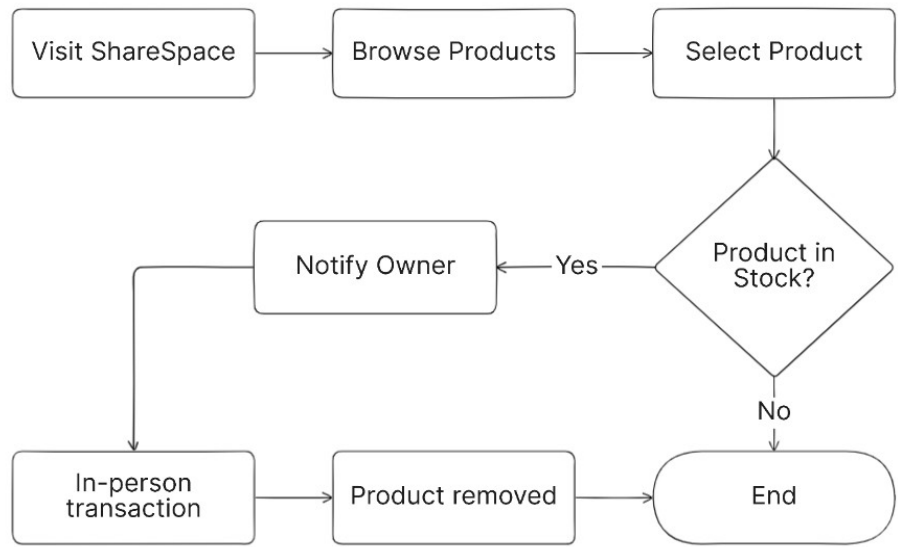

- Users search for their desired products from the search box through keyword matching, assisted by filters and category selections.

- Users can select products to view their full details and check users rating and credibility points.

- On finalizing, the user can request the product which provides them with the seller's details and also sends the seller a notification with the buyer's details. The product is also removed from the global pool of products and moved to a local pool of buyers requested products till the transaction is complete.

- Using each other's details, the buyer and seller can discuss and meet together in person to exchange the product and money.

- On every login after the user has received a product request, the user is prompted if the item is still in stock or has been sold.

- The user can either select sold, after which the product is deleted or pending which means the transaction has not taken place yet or declined which means the deal was called off the product once again joins the global product pool.

# 3. Challenges addressed

## Difficulty in moderation

This section details how ShareSpace solved the challenge of content moderation. Traditional systems implement complex keyword blacklisting approach. This leads to many false categorizations, leading to the need for manual moderation by human reviewers. This leads to high operational costs and is not sustainable in the long run.

ShareSpace solves this issue utilizing recent developments in generative AI and large corporations being able to train enormous GPTs (Generative Pre-trained Transformer). Multi-modal AI enable combining analysis of both image and text data best suited for e-commerce websites featuring both product images and descriptions.

ShareSpace utilizes Google Generative AI SDK to access their multi-modal AI model Gemini. The large well-trained AI models ensures images are well recognised and their advanced Natural Language analysis abilities ensure identification of attempts to disguise and masquerade non-compliant products. It prefaces the AI API call with a blacklisted keyword check to reduce API calls and response times.

**Functioning:**

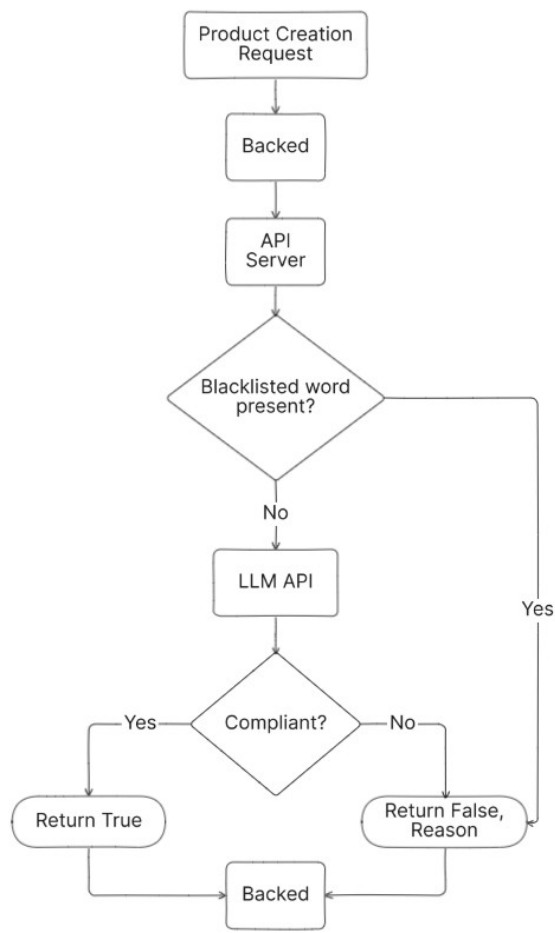

- The backed receives create product request from frontend.
- The request is forwarded to a Compliance checking API, which is responsible for checking the product listing's compliance with the platform's norms.
  The compliance check API first checks for presence of blacklisted words in the product text.
- If a blacklisted word is found, a false response is returned.

- If no blacklisted word is found, the request text and image are forwarded to the GPT API along with a system-prompt with guidelines and compliance check instructions.
- The API is forced to return a result in JSON format by a template with a compliance boolean and a reason if deemed non-compliant.
- The response from AI API is then returned by the compliance check API as well.
- If the backend receives the product as compliant, only then it adds the product to the database, else it is rejected.

## Incentivizing Economical Sharing

Almost all E-commerce platforms, whether made for everyone or specialised for specific communities, do not offer sellers any incentive to sell their products at economical prices. Lack of incentives lead to sellers prioritizing profit in the sale of their products rather than community sharing and helping out their peers.

ShareSpace solves this problem by introducing a Point based reputation system. Reputation points act as incentives for users to encourage positive behaviours and discourage negative behaviours.

### Reputation Point System:

Every user profile gets assigned a set number of points on registration. Points can be modified throughout the usage of the platform buy the user. Actions that lead to changes in the points linked to a user are called *modifiers*. Modifiers can be either positive or negative and have varying magnitudes. Positive modifiers lead to increment in points and include:

- Completing transactions
- Listing products for free
- Listing products at economical price

Negative Modifiers decrement points and include:

- Listing non-compliant products
- Performing activities in violation of platform's Terms of Service

In order to prevent point farming (amassing points by listing many fake products), points are credited to the user's account only after the product is sold.

### Point Incentives:
In order for reputation points to serve as incentives, they must provide some value to the user. ShareSpace points ensure users with higher points have their product listing appear higher in search results leading to higher sale chance. Higher points also increase user's reputation and trust among buyers.

## Competitive Price checking

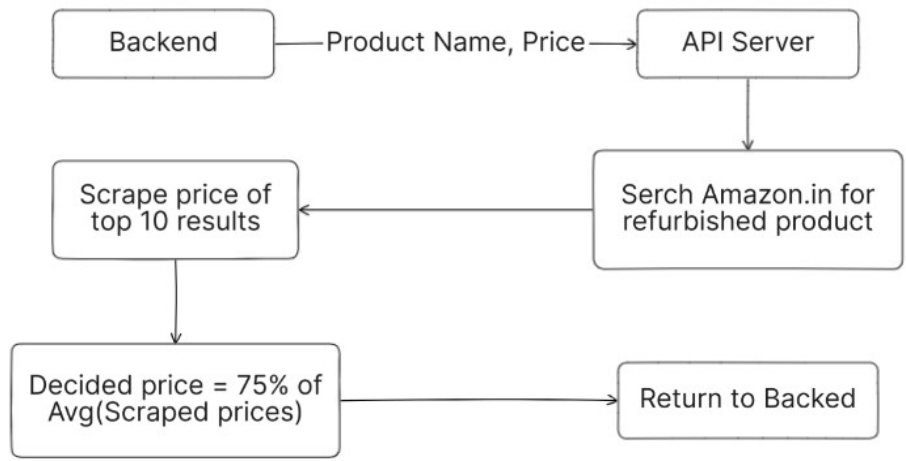

In order to incentivize users to list their products at economical prices, an approximate economical price for an item needs to be decided. This is especially difficult to do accurately as no definite price exists for second hand items. Yet ShareSpace aims to offer products more economically than E-commerce giants hence an easy way to determine an ideal cost would be to compare the prices against similar items on their websites.

ShareSpace solves this by searching the name of the product in the refurbished section on Amazon.in and takes the price of the top ten results and averages them. Then 75% of the average price is approximated as the ideal price limit for the product listing on ShareSpace. If the user mentioned price falls below the ideal price then the user is awarded a high amount of reputation points. If not then no points are awarded in relation to the price of the product.

# 4. Results

The development of ShareSpace has been successfully completed, achieving all the intended features and functionalities outlined during the design phase. A decent modern looking user interface has been achieved.

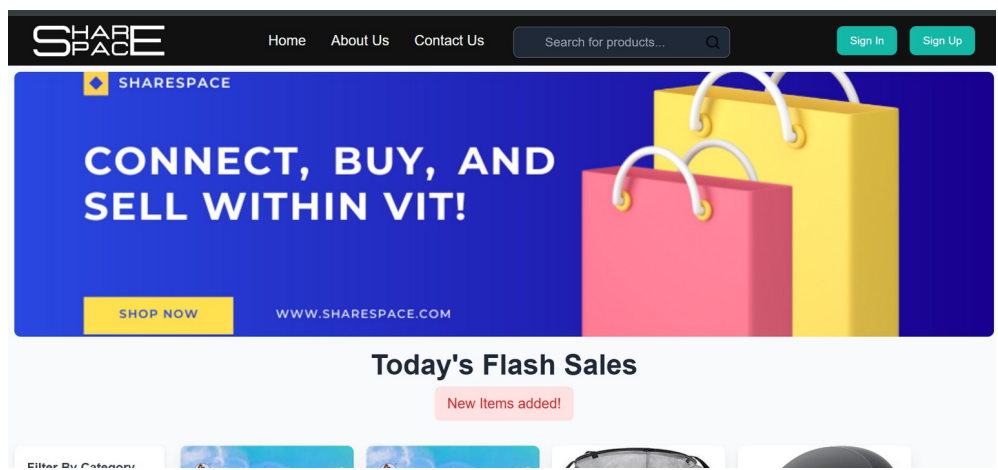

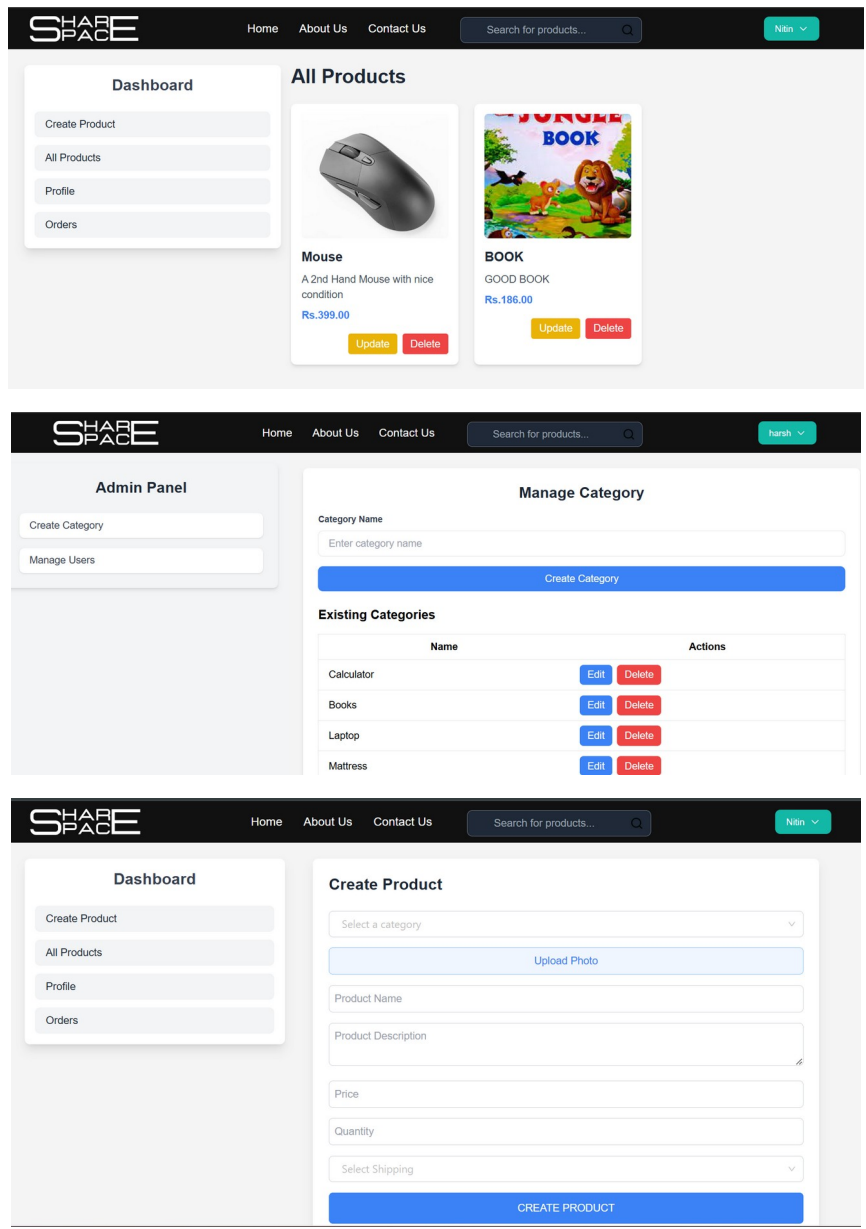

All the major features have been fully developed and are available on
https://github.com/nitin611/ShareSpace. All systems work as expected and the platform is
successfully able to handle many concurrent connections.

The Gemini AI integration for non-compliant product detection was also found to work flawlessly
during testing. The compliance check mechanism was tested with a diverse dataset of sample
products with varying levels of compliance, and the AI was able to distinguish them all
successfully. It maybe worked a bit too well, classifying all cases that could be considered on the
border of compliance as non-compliant.

The pricing validation system was also tested using various samples and provided an ideal cost for
almost all items that seemed economical enough for college students. All CRUD operations on
products and users were tested thoroughly and worked as expected.

## 5. Future Work

**Mobile Application**
To improve accessibility and user convenience, the development of a mobile application for ShareSpace is proposed. This application would offer identical functionalities to those available on the web platform, such as item listing and browsing, while being specifically optimized for mobile users. Utilizing cross-platform development frameworks like React Native or Flutter would guarantee compatibility across both Android and iOS devices.

**Training and hosting Custom AI modal locally**
Utilizing an AI modal trained specifically for the detection of non-compliant content or alternatively a fine-tuned general language modal might yield better results compared to employing a generic large language model. Furthermore, hosting the model locally within the backend infrastructure, as opposed to relying on external APIs for remotely hosted models, has the potential to decrease response latency.

**Expansion to other Universities**
Currently, ShareSpace is tailored for the author's university. Future iterations aim to expand the platform's applicability to other universities. Additionally, adjustments to verification processes and the establishment of rules and regulations tailored to individual colleges will be essential to ensure the platform remains localized and relevant. Collaboration with student organizations and university administrations will help promote the platform and address campus-specific needs.

**Introduction of Real-Time Chat**
Adding a secure real-time chat feature for buyers and sellers to communicate directly within the platform will prevent the need to use additional channels for communication and lead to more user privacy, as users may be unwilling to share their phone numbers or email ids for communication.

## Conclusion

In conclusion, ShareSpace addresses a critical need within student communities by providing a dedicated platform for sharing and selling second-hand goods. By leveraging modern technologies such as ReactJS, Node.js, MongoDB, and generative AI, the platform offers an adaptable, seamless, secure, and user-friendly experience tailored to the unique requirements of college campuses. ShareSpace not only bridges the gap between students with excess resources and those in need but also promotes sustainability by encouraging reuse and reducing waste. The integration of innovative features, such as compliance checks, competitive pricing validation, and reputation-based incentives, ensures that the platform aligns with its mission of affordability and community-driven sharing.

As an initial deployment at the author's university, ShareSpace has demonstrated significant adoption and impact, providing valuable insights into its potential scalability. Future enhancements, including a mobile application, locally hosted AI models, and expansion to other campuses, are expected to broaden its reach and effectiveness. By addressing the practical challenges, ShareSpace aims for a more sustainable and interconnected student community, serving as a model for similar initiatives in educational institutions worldwide.